**Bioinspired Materials with Self-Adaptable Mechanical Properties**

*Santiago Orrego[1,2,3], Zhezhi Chen,[1,2] Urszula Krekora[4], Decheng Hou[1,2], Seung-Yeol Jeon[1,2], Matthew Pittman[1], Carolina Montoya[3], Yun Chen[1], Sung Hoon Kang[1,2,5]\**

[1]Department of Mechanical Engineering, Johns Hopkins University, Baltimore, 21218, USA.
[2]Hopkins Extreme Materials Institute, Johns Hopkins University, Baltimore, 21218, USA.
[3]Department of Oral Health Sciences and Bioengineering Department, Temple University, Philadelphia, 19140, USA.
[4]Deparment of Chemical & Biomolecular Engineering, Johns Hopkins University, Baltimore, 21218, USA.
[5]Institute for NanoBioTechnology, Johns Hopkins University, Baltimore, 21218, USA.
E-mail: shkang@jhu.edu (S. H. Kang)

**Natural structural materials, such as bone and wood, can autonomously adapt their mechanical properties in response to loading to prevent failure. They smartly control the addition of material in locations of high stress by utilizing locally available resources guided by biological signals. On the contrary, synthetic structural materials have unchanging mechanical properties limiting their mechanical performance and service life. Here, a material system that autonomously adapts its mechanical properties in response to mechanical loading is reported inspired by the mineralization process of bone. It is observed that charges from piezoelectric scaffolds can induce mineralization from media with mineral ions. The material system adapts to mechanical loading by inducing mineral deposition in proportion to the magnitude of the loading and the resulting piezoelectric charges. Moreover, the mechanism allows a simple one-step route for making graded materials by controlling stress distribution along the scaffold. The findings can pave the way for a new class of self-adaptive materials that reinforce the region of high stress or induce deposition of minerals on the damaged areas from the increase in stress to prevent/mitigate failure. They can also contribute to addressing the current challenges of synthetic materials for load-bearing applications from self-adaptive capabilities.**

The mechanical efficiency of natural (or biological) structural materials is remarkable. Self-optimized architectures,[1] outstanding mechanical properties with weak constituents,[2] hierarchical structuring,[3] and self-adaptability of stiffness in response to external stimuli[4] are just a few examples of inspiring features that should be included within synthetic structural materials. They provide inspiration for potential solutions to address challenges associated with traditional approaches of material selection.[5] In the classical approach, a material is selected for a specific application based on the expected loading conditions, design objective and constraints, and databases of material properties with fixed values (e.g., Ashby method[6]). This methodology does not account for unpredictable loading conditions. To address this issue, safety factors are added, which increases the associated costs and weight of the structure; thus, reduces its mechanical efficiency.

A potential solution for this dichotomy is to have materials with self-adaptable properties responding to loading conditions. This feature can result in materials with improved mechanical efficiency and reduction in costs, resources and environmental impact. So far, there are few synthetic materials capable of increasing stiffness in response to external stimuli. For example, Capadona *et al.* and Ramirez *et al.* reported polymers that changed elastic moduli by changing the degree of crosslinking upon chemical, thermal or mechanical stimuli.[7,8] Agrawal *et al.* reported an increase of the elastic modulus of liquid crystal elastomers during cyclic mechanical stress due to the alignment of nematic directors.[9] Recently, a "self-growing" hydrogel responding to repetitive mechanical stress was reported through mechanochemical transduction.[10] Despite progress there are limited synthetic material systems adapting mechanical properties to external mechanical stimuli. Challenges of current self-stiffening materials include limited load-bearing capability, difficulty of material synthesis, high cost, lack of or limited biocompatibility, and need of additional energy for material property changes.[11]

Natural structural materials have resolved these challenges by taking different strategies.[12] Bone,[13] wood,[14] fish scales,[15] and coral reefs[16] are dynamic structural materials capable of self-adapting their mechanical properties in response to external loading to improve their mechanical efficiency and prevent failure.[13,17] Introducing these features to synthetic material systems has been challenging[18] since it requires active maintenance of living organisms. These natural materials utilize resources available in the environment to regulate their properties through mechanobiological mechanisms.[16,19] For example, bone and coral reefs utilize cellular signals to control the addition and removal of minerals harvested from surrounding media (e.g. blood or seawater) at specific locations.[20] During the process, organic matrices serve as templates for mineral growth. Specifically, it is reported that negatively charged carboxyl groups can bind calcium ions and induce nucleation of biominerals.[21] Inspired by these findings, there have been numerous reports[22] on mineral growth using negatively charged surfaces obtained by using electric field[23] or self-assembled monolayers.[24] However, previously synthesized minerals cannot change their mechanical properties upon external loading.

Here we report a synthetic material system that can change its mechanical behaviors in response to external loading conditions inspired by natural mineralization processes in bone and coral reefs. We hypothesized that if we have a scaffold that can generate charges proportional to an external mechanical stimulus, the charges can work as signals to induce mineralization from resources (e.g. mineral ions) available in surrounding media. As a result, the material system can exhibit self-adapting mechanical behaviors. Our material system uses a new mechanism to realize self-stiffening materials and overcomes limitations of current synthetic materials with changing mechanical properties from its enhanced load-bearing capability, low cost, simple synthesis, no need of extra energy, and biocompatibility.

To test our hypothesis, we utilized piezoelectric materials as scaffolds because they convert mechanical forces into electrical charges while other materials (e.g. triboelectric, flexoelectric materials) could be utilized. We immersed piezoelectric polymer films (PVDF, $(CH_2CF_2)_n$), $d_{33}$~30 pC/N) with oppositely charged surfaces in a simulated body fluid (SBF)[25] mimicking the ionic concentrations of human blood plasma (**Figure 1**a). After one week of SBF incubation, we observed the minerals formed on the surfaces of the films using a scanning electron microscope (SEM). Minerals were preferentially deposited onto the negatively charged surfaces (Figure 1b). This was consistent with previous study.[23] To characterize chemical compositions of the formed minerals, we conducted energy dispersive X-ray spectroscopy (EDX) and X-ray diffraction (XRD) measurements. EDX results showed peaks of Ca, P, and O, which are elements consisting of hydroxyapatite (HAp) (Figure S2). XRD analysis showed the characteristic diffraction peaks of HAp (Figure 1d). We also observed similar mineral deposition behavior on different substrates including piezoelectric composites (See Supporting Information S3 for details). Moreover, we found that we can control the type of formed minerals by controlling the ion composition of the media (See Supporting Information S4). Refreshing the SBF solution everyday led to the formation of HAp minerals (Figure 1d), whereas using non-refreshed SBF led to a combination of Calcite and HAp (Figure 1c). The minerals found in bone (HAp) and coral reefs (Calcite) originate from blood and seawater, respectively.[21] Metallic medical devices are typically coated with these type of minerals to enable biocompatibility.[26] Usually these coatings wear out and do not regenerate.

To understand the dynamics of mineralization, we conducted quantitative studies by comparing the mineral thickness at different conditions (**Figure 2**). First, we studied the effect of the polarity of the charged surface on the mineral formation rate (see Supporting Information S5). We found that the negatively charged surface showed about an order of magnitude faster mineral formation rate than the positively charged surface (Figure 2b). The mineral growth rate (~6.4 μm/day) was comparable to that reported by Yamashita *et al.*, which was obtained in polarized negative surfaces.[23] We also studied the effect of the concentration of the SBF on mineral formation rate. We found higher concentrations of calcium in SBF resulted in faster rates (Figure 2c), which is consistent with previous study that utilized other negatively functionalized surfaces.[27] In our study, we mainly utilized 10× calcium concentration in SBF as it provides faster mineral formation. We also found that we could modulate the mineral growth rate by controlling the external mechanical loading. Higher electrical charges generated by cyclic mechanical loads led to higher quantities of minerals compared to static or no mechanical loads (Figure 2d). Different from previous studies that utilized permanent chemical functionalization,[28] one can utilize "mechanical functionalization" using piezoelectric scaffolds to dynamically control the formation of minerals by modulating the loading condition.

As an application of the mineralization mechanism controlled by mechanical stress, we made a graded material by using a simple one-step process. A piezoelectric PVDF film was subjected to a cantilever loading since it offers a gradual increase of stress ($\sigma$) from the free end to the fixation point (see Figure 2e, Supporting Information S13). In this beam configuration, the electrical charge changes proportionally to the beam length. We experimentally confirmed higher electrical charges at regions of high stress and vice versa at regions of low stress (Figure 2e) as a piezoelectric charge ($Q$) is proportional to the applied stress ($Q = d\sigma A$, $d$: piezoelectric coefficient, $A$: area). Then, we submerged a PVDF specimen in SBF under repetitive loading (Supporting Information S7) and measured the amount of mineral along the beam (Supporting Information S8). Our results showed a gradual formation of minerals with higher amounts at regions of high stress (high charge) and vice versa at low stress regions (Figure 2f). This result indicates that our material system could autonomously reinforce regions that experience higher stress with more minerals. The mechanism allows materials or structures to self-adapt to the external loading conditions to minimize failure. We further investigated the correlation between the mechanical stress and the mineral distribution. We used a numerical model to calculate the stresses distribution along the beam and compared with the mineral thickness distribution. Interestingly, the stress distributions at different positions of the specimen showed similar trends as the mineral thickness (Figure 2f and Supporting Information S9). We also obtained the quantitative relation between the mineral height and the charge along the beam as Equation (1) (Supporting Information S13) and found that the mineral height (*MH*) is approximately proportional to the square root of the charge (*Q*) or stress (*σ*).

$$Q \cong d_{33}A(4.33 \times 10^4 MH^2) \tag{1}$$

(*Q*: charge in C, *MH*: mineral height in mm, *d*: piezoelectric coefficient in pC/N, *A*: area in mm$^2$).

Furthermore, we investigated whether our material system can change its mechanical behavior by inducing mineral formation in response to the external loading akin to bone (**Figure 3**a). We prepared porous piezoelectric scaffolds ($d_{33}$~2 pC/N) using electrospinning (Supporting Information S10) inspired by bone's porous architecture. The scaffolds were subjected to cyclic mechanical loading while submerged in SBF solution at room temperature (Supporting Information S11). The applied load and actuator displacement were measured to calculate changes in material property such as Young's modulus.

From the measurements, our material system showed self-stiffening behaviors by changing its mechanical properties in response to periodic loadings (Figure 3b). The modulus increased ~20% after 3 days of immersion in SBF. A control sample submerged in deionized water (no mineral ions available in the solution) did not show a significant increase in the mechanical properties over time (Figure 3c). Also, comparison of the stress-strain behaviors of the scaffolds before and after the external stimulation in SBF showed that the mechanical property increased ~30% in modulus and ~100% in toughness (Figure 3c). SEM images confirmed the mineralization of the scaffold that enabled the increase in mechanical property (Figure 3d-e). EDX results confirmed the presence of minerals with traces of calcium, phosphorous, and oxygen (Figure 3f). Fluorescent microscopy images showed the formation of minerals around the fibers (Figure 3i and Supporting Information S12).

Moreover, to further test the self-adaptive behavior of the material system, we subjected scaffolds to cyclic mechanical loading with different load magnitudes while submerged in SBF.

Before cyclic loading, we verified the different piezoelectric charges under different load magnitudes. We measured the stress-strain responses after samples were submerged in SBF. As expected, the modulus of the material system increased proportional to the magnitude of the load and the resulting piezoelectric charge (Figure 3g-h). Specifically, the modulus increased by ~180% after three days for the maximum 5N cyclic load, further demonstrating the self-adaptive capability of the material system.

In summary, we report a synthetic material system capable of changing its mechanical properties depending on external loading conditions by utilizing scaffolds that generate "signals" to induce mineralization from media with mineral ions. The material system allows control of the location and the rate of mineral formation by modulating the loading conditions. Our findings can contribute to addressing current challenges of synthetic materials used for load-bearing applications from self-adaptive capabilities. To make the material system portable, one can utilize porous scaffolds with suitable surface properties[29] for making liquid-infused porous piezoelectric scaffolds inspired by bone, which holds blood within porous matrices. Beyond biominerals used as a model system in this study, we can further expand the material system based on the reported mechanism for broader applications. We envision that the strategy and the mechanism that we report will open new opportunities for technological advancement including smart coatings for bone implants to mitigate mismatch of mechanical properties, scaffolds for accelerating treatment of bone-related disease or fracture, smart resins for dental treatments requiring tissue regeneration, and 4D materials and structures reconfiguring to adapt to loading conditions. They will also contribute to fundamental advances in understanding dynamic behaviors of adaptive materials and structures and dynamics of mineralization induced by mechano-electric coupling. Moreover, built upon recent studies that reported piezoelectric charges could stimulate cell differentiation in various cell types and improve cell adhesion, bone growth and function in endothelial, nerve, and bone cells,[30] the material system can serve as a model system for quantitative understanding of the process with adequate choice of medium and provide guides for facilitating the regeneration.

**Experimental Section**

*SBF Incubation:* Piezoelectric PVDF films (TE Connectivity, PN: 3-1003352-0) were submerged in a sealed polypropylene bottle (Dyn-A-Med, 80094) containing ~200 mL of SBF solution kept at 37 °C for the duration of the experiment (see Figure 1a). Every 24 hours, the sample was taken out of the beaker, gently rinsed with de-ionized water (Milli pore, Milli-D) and put back into a refreshed solution. After the incubation, the sample was repeatedly rinsed gently in a de-ionized water solution to remove salts of NaCl. The sample was taken out and dried at room temperature in air for further analysis including SEM, EDS, XRD, profilometer.

*Self-Stiffening Experiments*: To test the self-stiffening capability of our material system, electrospun PVDF samples were submerged in SBF solutions and were subjected to cyclic mechanical loadings to activate the piezoelectric charge generation. We prepared electrospun PVDF samples of 30×15 mm$^2$. Simulated body fluids (10×SBF) were prepared by following protocols described in Supporting Information S1. While submerged in the solution at room temperature, samples were subjected to cyclic compression loading using an electromechanical universal testing machine (TA Instruments Electroforce 5500 with a 10 N capacity load cell). The stiffness (*S*) was calculated from the load (*P*) and displacement ($\Delta\delta$) amplitudes measured with the actuator using the follow relation: $S=\Delta P/\Delta\delta$. The force and displacement signals were monitored and recorded during experiments. To prevent any change of mechanical properties of electrospun films due to viscoelastic and time-dependent behavior during experiments in SBF, we initially subjected samples to cyclic compression loading while submerged in DI water for 3 days. The modulus (*E*) was calculated by multiplying by the cross section area and the deflection $E=S*\Delta\delta / A$.

**Acknowledgements**
We would like to thank Mr. Eugene Kang for the help fabricating the electrospinning apparatus, Mr. Ian McLane for the help and guidance during contact profilometer measurements, and Mr. Prashant Ray for his help during the concentration measurements. This work is supported by the Air Force Office of Scientific Research Young Investigator Program Award (Award number: FA9550-18-1-0073, Program manager: Dr. Byung-Lip (Les) Lee), Johns Hopkins University Whiting School of Engineering start-up fund, and Temple University Maurice Kornberg School of Dentistry start-up fund (PI: Orrego). Any opinions, finding, and conclusions or recommendations expressed in this material are those of the author(s) and do not necessarily reflect the views of the United States Air Force.

**Conflict of interest**: A patent application has been filed on aspects of the described work.

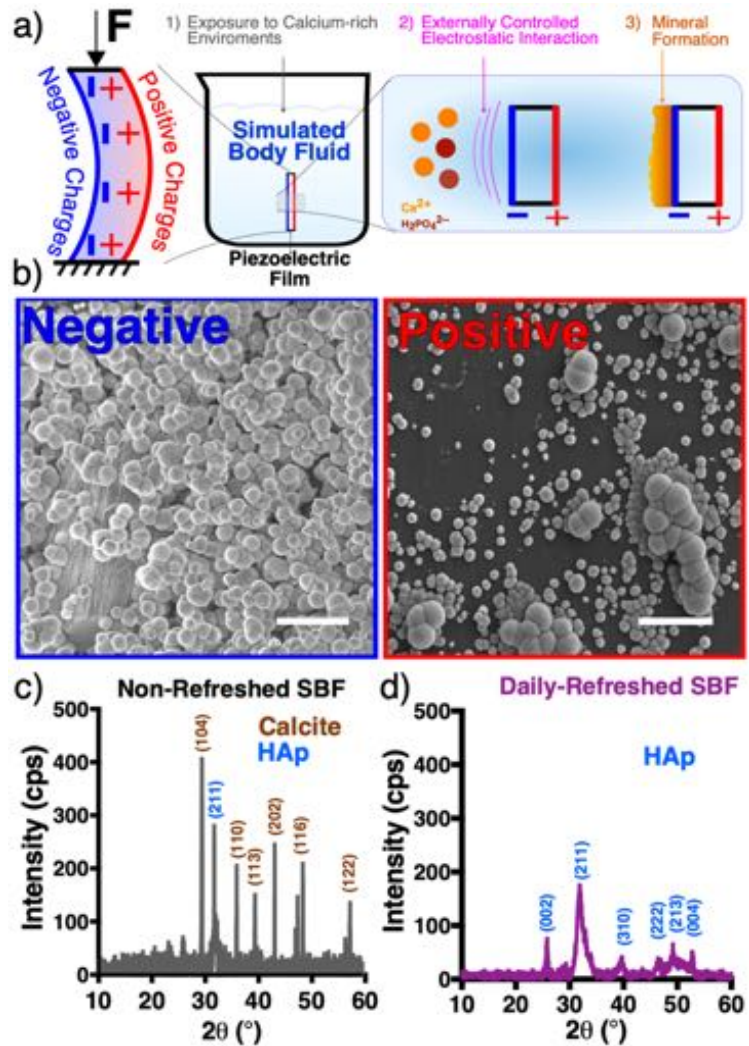

**Figure 1.** Formation of minerals induced by piezoelectric charges. a) Schematic of the experimental setup. A PVDF piezoelectric film ($d_{33}$~30 pC/N) was submerged in a simulated body fluid (SBF). The generation of negative charges controlled by external mechanical loading preferentially induced the formation of minerals on the negative side of the film. b) Micrographs of the negative (left) and positive (right) sides of the films showing significant formation of minerals on the negative side. c) The type of formed mineral was controlled by the medium used. Not refreshing the SBF solution everyday led to the formation of mainly Calcite minerals. d) Refreshing the SBF solution everyday led to the formation of mainly HAp minerals. Scale bars are 10 μm.

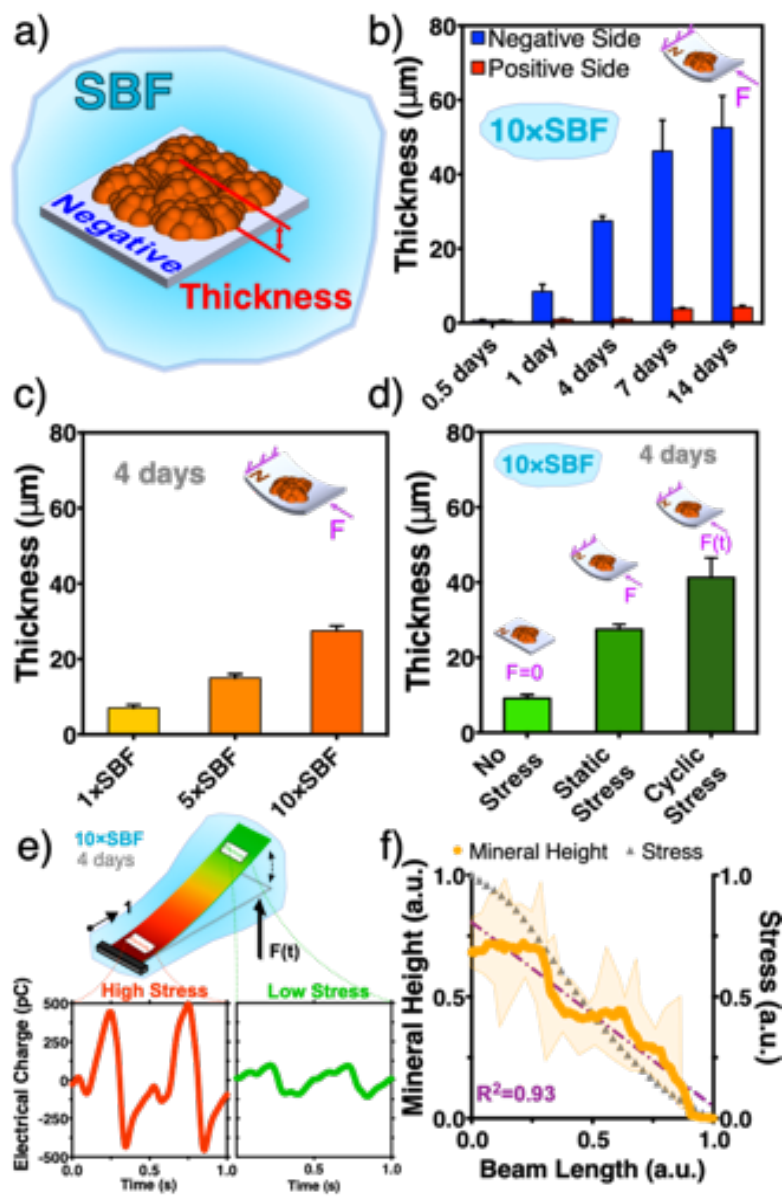

**Figure 2.** Dynamics of the formation of minerals by piezoelectric charges. a) The amount of mineral formed on films was quantified by measuring the mineral thickness. b) Time dependence test. A maximum of 60 μm of mineral was formed on the negative side after 2 weeks of incubation in SBF. No significant mineral was formed on the positive side. c) SBF concentration dependence test. Higher concentration of the SBF solution led to higher amounts of mineral. d) Mechanical loading dependence test. The higher charge generated by external loading led to higher amounts of mineral. e) Formation of graded materials. A piezoelectric film submerged in SBF and subjected to cantilever loading generated higher charge at regions of high stress (close to the foundation) compared to regions of low stress (low charge generation). f) Mineral was formed proportional to the mechanical stress (gray line) with approximately linear ($R^2=0.93$) distribution form the foundation to the free end (orange line). The orange line is the filtered (1D- median filter) data of the mineral height shown as shades in orange. The error bars were obtained by having one standard deviation from N=5 samples.

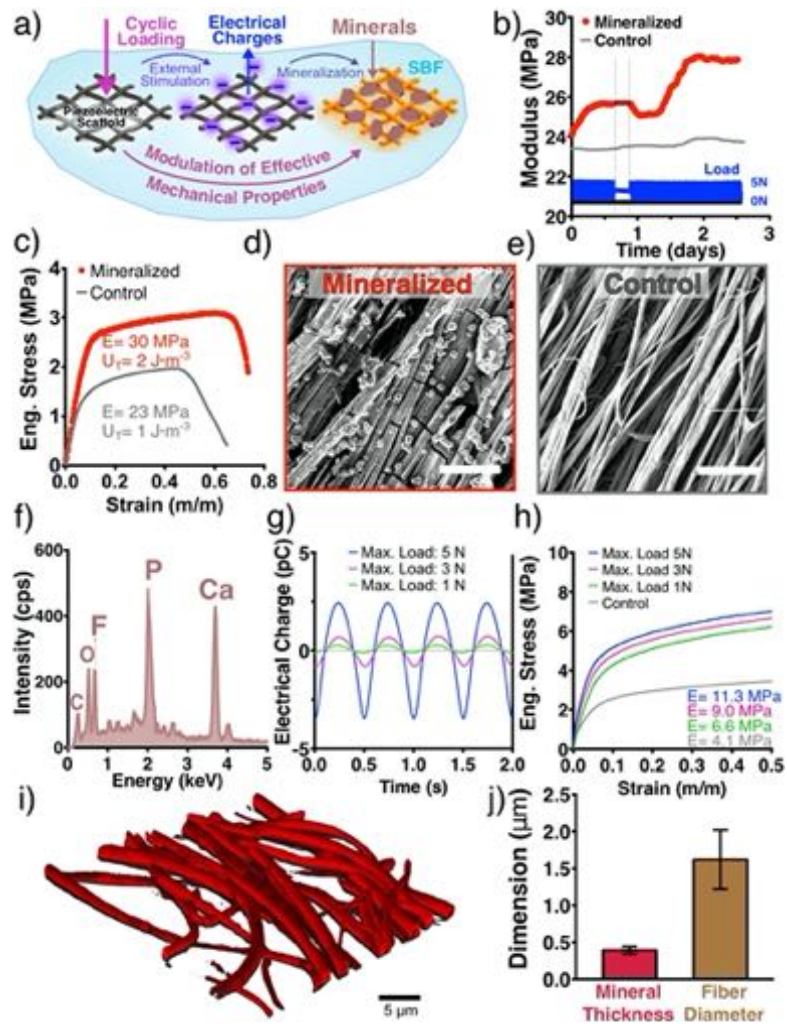

**Figure 3.** Self-adaptable mechanical properties. a) Electrospun piezoelectric scaffolds ($d_{33}$~ 2 pC/N) were submerged in SBF and subjected to controlled repetitive mechanical loading. The external activation of the electrical charges promoted the formation of minerals on the scaffold. The modulus of the mineralized scaffold was monitored for the duration of the experiment. b) The material showed a self-adapting mechanical property by increasing modulus under dynamic loading. A control sample submerged in deionized water (gray line) showed no increase of modulus as a function of time. c) After the repetitive loading experiment, samples were loaded-to-failure under quasi-static conditions to confirm the change of mechanical properties. The modulus (E) and toughness (U) increased ~30 and ~100%, respectively. d) Micrographs of the PVDF scaffold after submerged in SBF showing the formation of minerals, compared to e) the as-prepared scaffold. f) Chemical analysis (EDS) confirmed the formation of calcium phosphate minerals on the PVDF scaffold. g) To show the adaptability of mechanical properties, PVDF scaffolds were subjected to different magnitudes of mechanical stress and resulting electrical charge. h) The mechanical modulus increased proportional to the magnitude of the applied load during incubation. i) Confocal laser scanning microscopy showed the formation of minerals around the PVDF template. j) The average thickness of the formed minerals was 0.39 μm. The error bars were obtained by having one standard deviation from N=5 samples.

# Supporting Information

**Bioinspired Materials with Self-Adaptable Mmechanical Properties**

*Santiago Orrego, Zhezhi Chen, Urszula Krekora, Decheng Hou, Seung-Yeol Jeon, Matthew Pittman, Carolina Montoya, Yun Chen, Sung Hoon Kang\**

Table of Contents

**Supporting Information S1: Mineral Incubation Procedure**

Commercial piezoelectric PVDF films (TE Connectivity, PN: 3-1003352-0) were cut into 2×4 cm$^2$ specimens and hanged vertically in a sealed polypropylene bottle (Dyn-A-Med, 80094) containing ~200 mL of SBF solution (**Table S1**) kept at 37 °C for the duration of the experiment (see **Figure 1**a). Every 24 hours, the sample was taken out of the beaker, gently rinsed with de-ionized water (Milli pore, Milli-D) and put back into a refreshed solution. After the incubation, the sample was repeatedly rinsed gently in a de-ionized water solution to remove salts of NaCl. The salinity of the rinsing solution was measured by a salinity meter (Extech, EC170) and after reaching a magnitude below 0.1 ppt, the sample was taken out and dried at room temperature in air for further analysis.

To study the influence of ion concentration on mineralization induced by piezoelectric charges, different stable Calcium-saturated solutions were utilized including 1×,[1] 1.5×,[2] 5×[3] and 10×[4] SBF. The solutions were prepared as indicated by dissolving reagent-grade as received materials of NaCl, NaHCO$_3$, KCl, K$_2$HPO$_4$·3H$_2$O, MgCl$_2$·6H$_2$O, CaCl$_2$ and Na$_2$SO$_4$ (All from Sigma-Aldrich) in the deionized water in accordance with the following ion concentrations (Table S1). 1×SBF and 1.5×SBF are further buffered with trishydroxymethylaminomethane (Tris) and HCl until pH=7.4 at 37 ºC in an incubator as indicated in the recipe.

**Table S1**. Ion concentrations of 1×, 1.5×, 5× and 10× SBF Solutions (mM).

|  | Na$^+$ | K$^+$ | Ca$^{2+}$ | Mg$^{2+}$ | Cl$^-$ | HCO$_3^-$ | HPO$_4^{2-}$ | SO$_4^{2-}$ |
|---|---|---|---|---|---|---|---|---|
| 1×SBF | 142.0 | 5.0 | 2.5 | 1.5 | 148.8 | 4.2 | 1.0 | 0.5 |
| 1.5×SBF | 142.0 | 5.0 | 3.75 | 1.5 | 148.8 | 4.2 | 1.5 | 0.5 |
| 5×SBF | 714.8 | - | 12.5 | 7.5 | 723.8 | 21 | 5 | - |
| 10×SBF | 1030.0 | 5.0 | 25.0 | 5.0 | 1065.0 | 10.0 | 10.0 | - |

The change in pH of the SBF solutions was monitored. A pH meter (Extech PH220-C) was calibrated using a 3-point method including acid (pH=4.01), neutral (pH=7.00) and alkaline (pH=10.00) buffers. The probe was dipped in the solution and the pH value recorded at different time durations. **Figure S1** shows the change of a 10×SBF solution during incubation process of a PVDF piezoelectric film.

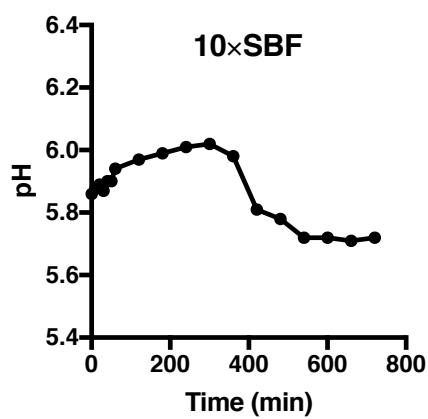

**Figure S1.** Change of pH for a 10×SBF solution kept at 36 °C during the incubation of minerals on a piezoelectric PVDF film for 12 hours.

**Supporting Information S2: SEM, EDS and XRD Evaluations**

A scanning electronic microscope (Tescan, Mira3) was used to observe the morphology of the formed minerals at an accelerating voltage of 10 kV. **Figure S2**a shows a representative layer of minerals formed on a PVDF film after 7 days of incubation in 10×SBF. Chemical and structural evaluations of the newly formed minerals were conducted by energy dispersion of X-ray (EDS; EDAX Co., USA, Octane Plus) analysis and powder X-ray diffraction (XRD; Philips, X'Pert Pro). Figure S2b shows the chemical evaluation of the mineral layer showed in Figure S2a. Clear peaks of calcium and phosphorus were observed with a ratio of Ca/P=1.47. Then, mineral powders were scratched off from a dried sample using a razor blade for XRD. XRD measurements were conducted with 10°-65° incident angle and step size of 0.02° using CuK$_\alpha$ radiation. A typical response of XRD results is shown in Figure S2c including a comparison with standard hydroxyapatite (HAp) readings,[5] which showed good agreement between peaks of formed minerals and those from standard HAp.

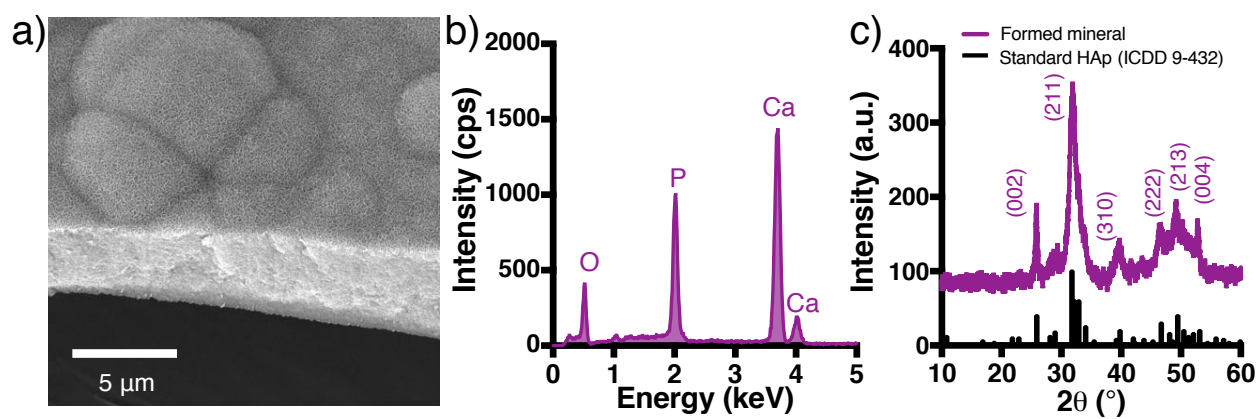

**Figure S2.** Mineralization of a piezoelectric PVDF film. a) SEM image of minerals formed on a negative side after 7-day incubation in 10×SBF with a 30° tilt view. b) Chemical evaluation of the mineral layer in the SEM image by EDS. c) XRD measurement of minerals from the specimen shown in the SEM image comparing with minerals formed on a PVDF film from the ICDD database.

**Supporting Information S3: Mineralization on Porous Piezocomposites**

Piezoelectric composites were utilized to check whether mineralizations induced by piezoelectric charges occur in a different material system. We employed a porous piezoelectric composite comprised of a PDMS matrix, carbon nanotubes and $BaTiO_3$ fillers. The sample was prepared by utilizing a combination of the fabrication methods described in the work by Li *et al.* [6] and Chen *et al.*[7] Briefly, the piezocomposite samples were prepared via solvent-casting by thoroughly mixing PDMS (base material:crosslinking agent = 10:1 by weight) (Sylgard 184, Dow Chemical) with $BaTiO_3$ nanoparticles (diameter: ~200 nm, US3830, US Research Nanomaterials, Inc.) as 15 wt% of PDMS, carbon nanotubes (OD: 10-30 nm, US4500, US Research Nanomaterials, Inc.) as 1 wt% of PDMS and sacrificial NaCl microparticles (Sigma Aldrich S7653) as 26.5 wt% of PDMS. To thoroughly mix the compounds, a revolutionary mixer (KK-400W, Mazerustar) was utilized for 360 seconds. The mass ratio between $BaTiO_3$,

NaCl and PDMS was adjusted to create polymer composites with different porosities. The mixture was then cured in an oven (Lindberg Blue M, Thermo Fisher Scientific) at 60 °C for 6 hours. After 24 hours, the solidified composite was immersed in water overnight to completely remove the NaCl grains. Microscale pores were then created. The synthesized microporous polymer composite was then cut with a razor blade into any shape and size according to the experimental requirements. **Figure S3**a shows a representative sample at the macroscale. The sample is soft and can allow high bending curvature without failure. The microstructure of a pristine sample is shown in Figure S3b. The average pore size is ~400 μm diameter. To improve the elcetromechanical output, samples were subjected to corona poling for 3 hours at 20 kV and 120 °C.[6] The measured piezoelectric coefficient of the porous piezocomposite is $d_{33}$~30 pC/N and obtained as explained in Supporting Information S6.

After fabrication, the piezocomposite was incubated in 10×SBF for 7 days under cyclic mechanical loading as explained in Supporting Information S1. After incubation, the sample was cut in half using a razor blade, and the fractured surface was evaluated in SEM and EDS as explained in Supporting Information S2. Figure S3c shows the fracture surface on the center of the sample after 7-day incubation under cyclic mechanical stimulation. The formation of the minerals inside the pores can be observed, which are denoted with white arrows. Chemical analysis of these sample area confirmed the presence of calcium phosphate minerals and peaks of Si corresponding to the PDMS matrix (Figure S3e). A higher magnification view in one of the pores can clearly show the minerals (see Figure S3d). The corresponding chemical analysis confirmed the presence of calcium and phosphorus corresponding to the ions available in the SBF solution (Figure S3f).

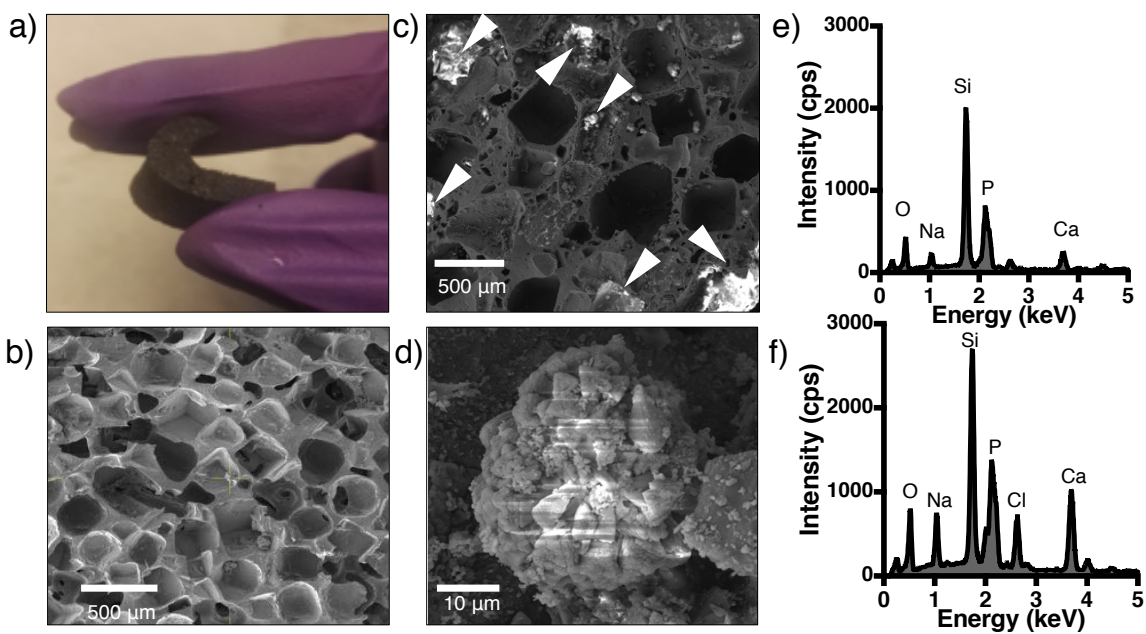

**Figure S3.** Mineralization of porous piezoelectric composites. a) Macroscale view of the sample. b) SEM image of the cross-section of the as-prepared porous piezoelectric composite showing porous microstructure. c) Cross-section view of the porous microstructure after 7-day incubation in 10×SBF and under cyclic mechanical loading showing mineral formation inside the pores as marked by white arrows. d) Higher magnification view of the mineral. e) Chemical (EDS) evaluation of the section view in c) showing peaks of calcium, phosphorus and oxygen corresponding to the newly formed minerals, and peaks of Si corresponding to the PDMS matrix. f) Chemical (EDS) analysis of the section view in d) showing peaks of calcium, phosphorus and oxygen corresponding to minerals and peaks of Si corresponding to the PDMS matrix.

**Supporting Information S4: Ion Concentration Measurements of SBF**

To study the potential change of calcium and phosphate ions present in SBF over time as a result of mineral formation, we conducted ion concentration studies in the solutions. Calcium concentration in the SBF solution was measured using the protocol outlined by Gindler and King.[8] First, stock solutions of dye and base were prepared. The stock dye solution was prepared by combining 0.018 grams of methylthymol blue sodium salt (CAS 1945-77-3), 0.720 grams of 8-Hydroxyquinoline (Sigma Aldrich, CAS 148-24-3), and 1.00 mL of 12M HCl (CAS 7647-01-0) in 100 mL in deionized water. The stock base solution was prepared by combining 2.40 g of sodium sulfite (Fisher Scientific, CAS 7757-83-7) and 22.0 mL of monoethanolamine (Sigma Aldrich, CAS 7757-83-7) in 100 mL of deionized water. Working reagent was made by mixing equal amounts of stock dye solution and stock base solution.

First, calibration measurements were done using the solutions containing known concentration of calcium. For each solution with known concentrations of calcium, 0.050 mL was mixed with 3.00 mL of working solution inside of a cuvette (Sigma Aldrich C5677). Using a UV-Vis photo spectrometer (LAMBDA 950, Perkin Elmer), absorbance values were measured from each sample. After confirming that the absorbance peak was at or near 612 nm, the absorbance reading was taken for each sample. After measuring absorbances of solutions with known concentrations treated with this assay, the relationship between calcium concentration and absorbance at 612 nm was found and described in Equation S1):

$$C = \frac{A - 0.244}{0.013} \quad \text{(S1)}$$

where $C$ is calcium concentration in mmol/L and $A$ is absorbance. This relationship was used to estimate the calcium concentration values in SBF solutions from absorbance readings.

To test calcium concentrations in SBF buffers, 0.050 mL of SBF solution was mixed with 3.00 mL of working solution in a cuvette. Calcium concentrations were correlated to absorbance reading using the calibration curve and Equation S1. The change of calcium concentration in an SBF solution is presented in **Figure S4**.

Phosphate concentrations in solution were measured using the protocol outlined by Karl and Tien.[9] Stock potassium antimonyl tartrate solution was prepared by dissolving 0.2743 g of potassium antimonyl tartrate (Sigma Aldrich CAS 331753-56-1) in 100 mL of deionized water. Mixed reagent was prepared by combining 25 mL of 2.5M $H_2SO_4$ (Alfa Aeser CAS 7664-93-9) 0.300 grams of ammonium molybdate (Sigma Aldrich CAS 13106-76-8), 0.264 grams of ascorbic acid (Alfa Aesar CAS 50-81-7), and 2.5 mL of potassium antimonyl tartrate solution in a 50 mL volumetric flask and filling the remainder with deionized water and mixing by shaking until all solids dissolved.

Similar to the case of the calcium ion concentration measurement, a calibration curve was obtained by using solutions containing known concentrations of phosphate ions. For each sample with known concentrations of phosphate, 4 mL of sample, 800 μL of mixed reagent, and 200 μL of deionized water were mixed in a cuvette. Using a UV-Vis photo spectrometer, absorbance readings were taken from each sample. For each sample with unknown concentration of phosphate, 4 mL of sample, 800 μL of mixed reagent, and 200 μL of deionized water was mixed in a cuvette. The absorbance readings were taken at 840 nm. Phosphate concentrations were correlated to absorbance reading using the calibration curve. After measuring absorbances of solutions with known concentrations treated with this assay, the relationship between calcium concentration and absorbance at 840 nm was found to be Equation S2.

$$P = \frac{A+0.022}{0.252} \tag{S2}$$

where $P$ is Phosphate concentration in mmol/L and $A$ is absorbance. This relationship was used to calculate the phosphate concentration values of unknown solutions of SBF from absorbance readings. The change of phosphate concentration in an SBF solution is presented in Figure S4.

Error was calculated using the following formula:[10]

$$S_C = \sqrt{(\frac{\partial C}{\partial A})^2 S_A^2} \tag{S3}$$

where $S_C$ is the calculated error in concentration measurements and $S_A$ is the standard deviation of the absorbance readings.

For calcium concentration calculations, this simplified to Equation S4.

$$S_C = \sqrt{\frac{S_A^2}{0.013^2}} \tag{S4}$$

For phosphate concentration calculations, this simplified to Equation S5.

$$S_C = \sqrt{\frac{S_A^2}{0.252^2}} \quad (S5)$$

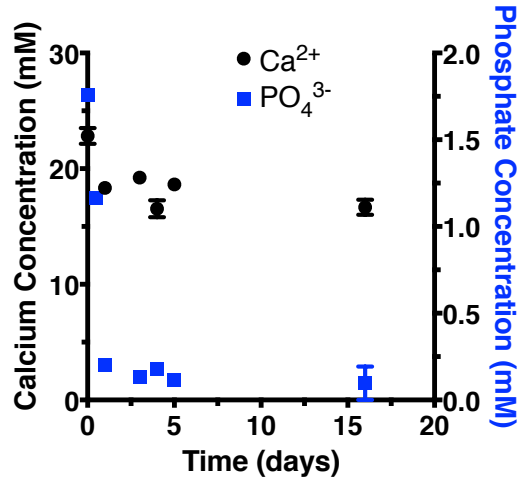

**Figure S4.** Change in concentration of calcium and phosphate ions over extended period of time in an SBF solution without being refreshed and maintained at 36 °C. The error bars are obtained by having one standard deviation from N=5 samples.

**Supporting Information S5: Mineral Thickness Measurements**

We utilized a contact stylus profilometer to quantify the height (thickness) of the new mineral layer formed on the piezoelectric films of PVDF. Minerals formed on the films in 10×SBF for different duration of times and mechanical loading conditions (see Supporting Information S1) were measured. The external edges of the sample were masked with Scotch® tape (7510000822520). After incubation period, the tape was gently de-bonded exposing the PVDF film. A step between the piezoelectric substrate and the newly formed mineral was created. The stylus of surface profilometer was placed on the mineral phase. Thickness measurements were performed on the mineral layer deposited on PVDF using the Dektak 3 surface profilometer with a 0.8 mm cutoff length according to ANSI standard B46.1. The step-height was measured by calculating the difference between the peak and valley ($R_y$) of the surface profile. Several profiles were measured along different locations of the mineral layer to calculate an average step-height per sample. The time-dependent thickness measurements are summarized in Figure 2b. The similar method was used for the SBF concentration dependence study as summarized in Figure 2c.

To show the influence of mechanical loading on mineralization rates, we conducted incubation in 10×SBF subjecting PVDF films to three different configurations including flat (no stress), fully rolled under static loading condition, and cantilevered bending under cyclic loading (see Supporting Information S7). We measured the mineral thickness for each configuration using the surface profilometer by following the aforementioned procedure as summarized in Figure 2d.

**Supporting Information S6: Piezoelectric Coefficient Measurements**

We measured the electromechanical response of the materials utilized in this investigation (i.e. commercial PVDF film, electrospun PVDF and piezoelectric composite) based on the Berlincourt's method.[11] Piezoelectric samples were sandwiched between compression platens covered with conductive carbon tape used as electrodes. The electrodes were wired to a charge amplifier (Piezo Film Lab Amplifier P/N 1007214, Measurement Specialties, Virginia). The amplifier was operated in Charge-Mode (100 pF) and bandwidth filter configured at 1 Hz and 10 Hz. Samples were subjected to cyclic compression loads (P) using an electromechanical universal testing machine (TA Instruments ElectroForce 5500). The load amplitude and frequency, and preload were adjusted to evaluate the electromechanical performance under different load conditions. The load cell and the charge amplifier signals were connected to a data acquisition system (NI-PCI 6251 and BNC-2110, National Instruments, Texas). Data was acquired with a sample rate of 1000 Hz ensuring our measures generated a sufficient representation of the acquired load and electrical charge signals. A schematic showing the wiring and configuration of the electromechanical measurements is showed in **Figure S5**a.

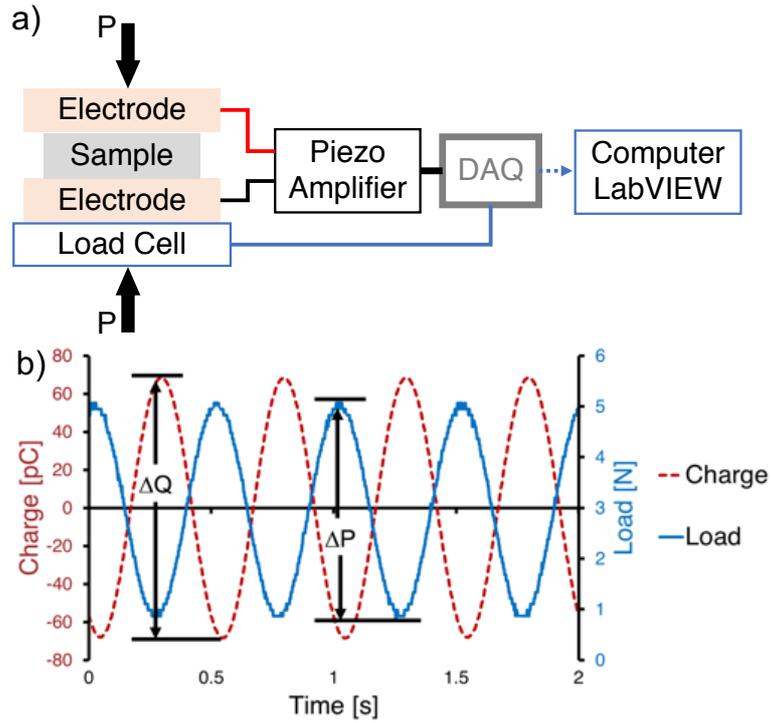

**Figure S5.** Piezoelectric coefficient measurements. a) Schematic of the configuration used to measure the electromechanical behavior of piezoelectric materials. Samples were sandwiched between two electrodes, which were wired to an amplifier and a data acquisition (DAQ). Signals were recorded and analyzed in LabVIEW. b) Electromechanical response of a commercial PVDF film. Signals recorded from mechanical load (blue) and electrical charge (red).

LabVIEW software was used to process the measured signals and calculate the piezoelectric properties. The peak/valley function was used to calculate the electrical charge (ΔQ) and load cell (ΔP) amplitudes. The piezoelectric coefficient was calculated as Equation S6.[12]

$$d_{33} = \frac{\Delta Q}{\Delta P} \tag{S6}$$

To validate electromechanical experiments, we used the commercial PVDF film (TE Connectivity, PN: 3-1003352-0) with known piezoelectric coefficient ($d_{33} \sim$ 30 pC/N) before each measurement (see Figure S5b).

**Supporting Information S7: Graded Material**

To investigate the potential capability of our material system to deposit different amount of minerals along a surface proportional to the amount of stress, we subjected piezoelectric films of PVDF to cantilever loading. With this loading configuration, regions of the film closer to the foundation experience high mechanical stress, thus high charge generation. Regions of the film closer to the point of load application will have low stress, thus low charge generation. This configuration can lead to having a graded material using a simple one-step procedure by generating graded stress distribution. We cut PVDF films of 5×40 mm$^2$ with a razor blade. One side of the film was fixed to a foundation. The other side was bent by an actuator excited by a long stroke shaker (Electro-Seis APS 113, APS Dynamics, California). Films were subjected to strain-controlled actuation with a stroke of ~25 mm, guaranteeing that samples were sufficiently bent. The electrical charges were measured in the regions of high and low stresses by carbon tape electrodes of ~3×5 mm$^2$. Electromechanical characterization was conducted as explained in Supporting Information S6. A schematic of the loading and charge measurement configuration is shown in **Figure S6**a. To allow mineral formation in response to the applied stress, the samples were submerged in a 10×SBF solution refreshed every day for 3 days under cyclic loading. After the incubation, samples were rinsed and prepared for surface profile measurements (Supporting Information S8).

**Supporting Information S8: 3D Optical Profilometer Measurements**

To measure the mineral distribution that was deposited on the PVDF films from the graded material experiments (Supporting Information S7), we utilized an optical profilometer. Samples were carefully attached to a silicon wafer by fixing with tape. A non-contact profilometer (Laser Scanning Microscope, Keyence VK-X100, Osaka, Japan) was used to scan samples over mineralized regions. The profilometer collected 3D information (x, y, z) from the scanned area. Samples were masked in their borders and a 3D distribution of the mineral was obtained. Several images were taken along the length of the sample and neighboring images were stitched to obtain the final result. First, scanned images were initially processed by VK analysis application (Keyence, Osaka, Japan) to correct surface tilting by checking the level of the substrate layer. Common correction methods we used were the 2-point linear profile correction and the 3-point non-linear profile correction, depending on specific situations. Data was then imported to Gwyddion (Department of Nanometrology, Czech Metrology Institute). We utilized analysis tools to quantify height distribution of minerals along the length of sample. Specifically, we obtained the mineral height along the beam length by calculating the root mean square mineral height on the film surface. The mineral height vs. beam length data was then filtered using a 1-D median filtering of order 300. This allowed the mineral distribution to be less noisy.

**Supporting Information S9: Numerical Model**

A three-dimensional (3D) finite element model was developed to simulate piezoelectric beam deflection. The simulation was conducted using the commercial FEA package Abaqus (SIMULIA, Providence, RI). The Abaqus/Standard solver was employed and the model was built using linear solid elements (Element type: C3D8). The beam geometry was set as 15 mm (length) × 0.1 mm (thickness) × 5 mm (width). The material was assumed to be linearly elastic, with Young's modulus of 4 GPa and Poisson's ratio of 0.18. The one end of the beam was constrained in all degree of freedom and the bottom surface of the beam was loaded with 500 Pa in the vertical direction of the beam (**Figure S6**). The beam deflection was 6.5 mm and stress ($S_{33}$) values on the top surface were extracted along the beam direction.

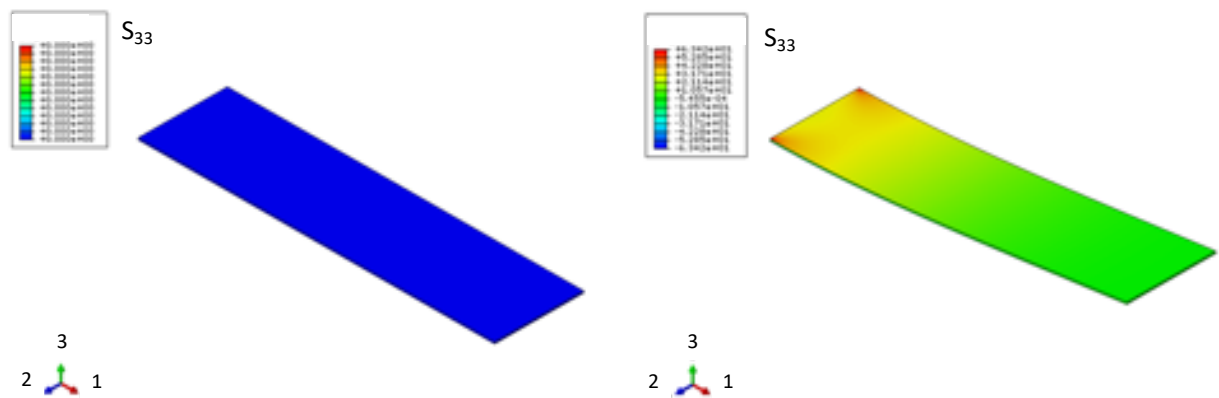

**Figure S6.** Finite element analysis (FEA) to calculate $S_{33}$ according to the piezoelectric beam deflection. Left image shows stress map ($S_{33}$) before deformation. Right image shows stress distribution ($S_{33}$) of beam after subjected to cantilever bending mode.

**Supporting Information S10: Electrospinning of Polyvinylidene fluoride (PVDF)**

To make porous piezoelectric scaffolds, electrospinning was utilized. PVDF pellets (Mw=530,000), N,N-dimethylformamide (DMF, 99.8%) and acetone (99.5%) were purchased from Sigma-Aldrich (St. Louis, MO, USA). All the materials were used as received without further purification. The PVDF pellets were first dissolved in a DMF/acetone mixed solvent at a certain concentration and stirred at 80°C overnight until a transparent homogenous solution was obtained. The weight ratio of the DMF and acetone in the mixed solvent was 2:3 and the mass concentration of PVDF in the solution was 18%. Then, the PVDF/DMF/acetone solution was cooled down to the room temperature.

To electrospin fibrous PVDF membranes, the PVDF/DMF/acetone solution was placed into a 10-mL glass syringe installed with a stainless steel-needle as the spinneret. A digitally controlled syringe pump (N1000 - New Era Pump Systems, Inc.) was used to feed the polymer solution into the needle tip at a constant feeding rate of 1 mL/hr. A rotary disk with the diameter of 30 cm was electrically connected to the ground and used as the fiber collector. The whole setup is shown in **Figure S7**a. During the electrospinning process, the PVDF solution underwent a high DC electrical field of 138 kV/m. Such a high DC electrical field was generated by applying a positive voltage of 18 kV with a 13 cm gap between the spinneret and the fiber collector. After electrospinning, a non-woven PVDF fibrous mat was formed onto the rotary aluminum disk, which was peeled off as a thin fibrous PVDF membrane (Figure S7b). The thickness of the PVDF members was controlled around 100 - 200 μm by the electrospinning time, which was around 2 hours. The surface morphology of the as-prepared electrospun fibrous PVDF membranes (Figure S7c) were characterized by using a scanning electron microscope (SEM, Tescan Mira 3). Prior to SEM examination of the fibrous membranes, the membrane specimens were sputter-coated with Platinum to avoid possible charge accumulation onto the PVDF fibers during the test.

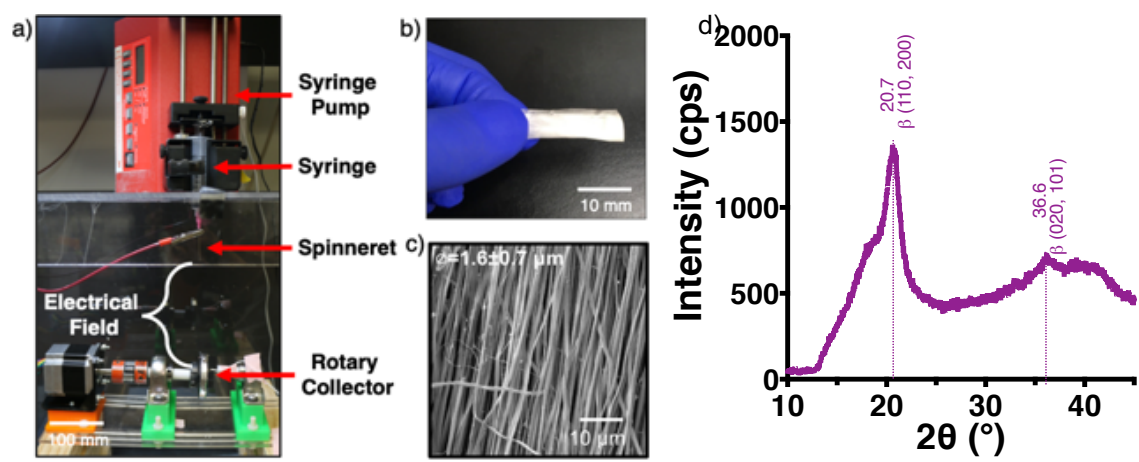

**Figure S7.** Electrospinning of piezoelectric PVDF. a) Apparatus utilized to conduct electrospinning. b) A photo of an electrospun PVDF membrane. c) SEM of the microstructure of PVDF. d) XRD pattern of a pristine PVDF electrospun sample.

The degree of crystallinity of the PVDF electrospun films (i.e. amount of β-phase) was estimated using XRD as described in Supporting Information 2. Figure S7d shows the XRD pattern of an electropun sample. The peaks observed 20.7°, 36.6°, and 56.9° are assigned to (1 1 0, 2 0 0) and (0 2 0, 1 0 1) diffractions of the β-PVDF crystal plane, respectively.[13] This phase has a large spontaneous polarization within crystalline phase thus providing the piezoelectric effect.[14]

**Supporting Information S11: Self-Stiffening Experiments**

To test the self-stiffening capability of our material system, electrospun PVDF samples were submerged in SBF solutions and were subjected to cyclic mechanical loading (i.e. external stimulus) to activate the electrical charge generation and promote the nucleation and growth of minerals that can modulate the effective stiffness by adding minerals. We prepared electrospun PVDF samples of 30×15 mm$^2$ according to Supporting Information S10. Simulated body fluids (10×SBF) were prepared by following protocols described in Supporting Information S1. While submerged in the solution at room temperature, samples were subjected to cyclic compression loading using an electromechanical universal testing machine (TA Instruments Electroforce 5500 with a 10 N capacity load cell). The stiffness ($S$) was calculated from the load ($P$) and displacement ($\delta$) amplitudes measured with the LVDT actuator using the follow relation: $S=\Delta P/\Delta \delta$. The force and displacement signals were monitored and recorded during experiments using Wintest software 7.6. To prevent any change of mechanical properties of electrospun films due to viscoelastic and time-dependent behavior during experiments in SBF, we initially subjected samples to cyclic compression loading while submerged in DI water for 3 days. The modulus ($E$) was then calculated by multiplying by the cross section area and the deflection $E = S\Delta\delta/A$.

After the modulus reached a constant value (**Figure** S8a), we switched the solution to the SBF and kept the same loading conditions. The force and displacement were continually monitored and the modulus was calculated based on the measured values as summarized in Figure 3b. To explore the modulus adaptation in response to the external mechanical loading, we subjected samples to different loading configurations in terms of frequency and load amplitude. We chose loading conditions that offered a variety of charge conditions whose results are summarized in Figure 3g.

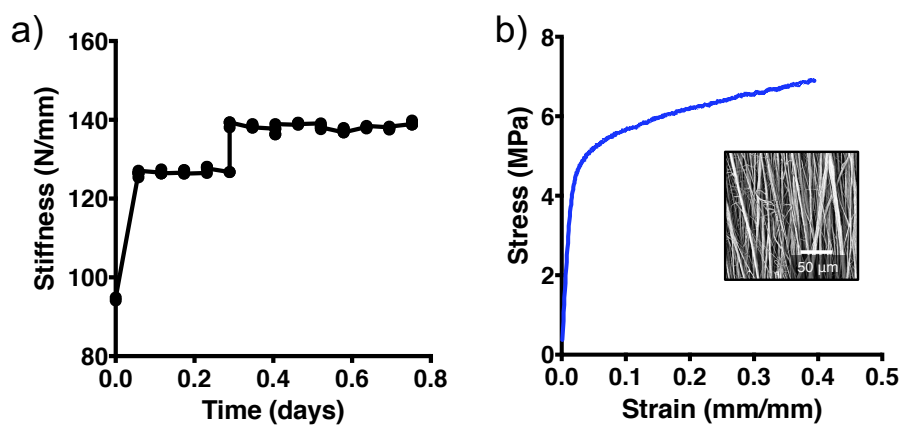

**Figure S8**. Preparation of self-stiffening experiments. a) Stiffness change of a PVDF electrospun sample submerged in water. Initial response after cyclic loading was recorded. Due to viscoelastic effects, stiffness reached a plateau after 12 hours. b) Stress-strain response of as-prepared electrospun PVDF specimen without any incubation or mineralization. The fibers were aligned in the direction of loading.

For measuring stress-strain responses of electrospun PVDF specimens, the specimens were subjected to quasi-static loading to failure under a tensile configuration using a universal testing system (TA ElectroForce 5500) with rate = 0.6 min$^{-1}$ based on the ASTM standard.[15] The tensile strength (σ) of the specimens was calculated according to σ = *P*/*A*, where *P* is the measured load and *A* the cross-sectional area (width by thickness) of the specimen. The strain was calculated by measuring the gauge length divided by the displacement amplitude recorded by the actuator. Figure S8b shows the stress-strain response of a representative sample of electrospun PVDF. The elastic modulus of the sample is ~20 MPa and the tensile strength ~4 MPa, which are comparable with values found in literature.[16] The stress-strain responses after cyclic loadings are shown in Figure 3c and 3 h.

**Supporting Information S12: Confocal Microscopy of Mineralized Electrospun Scaffolds**

Mineralized electrospun PVDF sample (7 days in 10×SBF, daily refreshed) was cut into 0.5×0.5 cm$^2$ piece and stained with 1 mg/mL Alizarin Red S (Sigma–Aldrich A5533) in aqueous buffer for 20 min at room temperature. After staining, the sample was gently rinsed with de-ionized water for 2 min and then kept in PBS solution (Lonza 17-516F) for imaging purpose. Imaging of mineral deposition was performed by confocal laser scanning microscopy on an inverted microscope system (Leica TCS SP8) with a 63x1.40 N.A. Plan Apo oil immersion objective lens. Illumination of the samples stained with Alizarin Red S was provided by a white light laser (Leica WLL), exciting at 540 nm. Images were captured by a hybrid detector (Leica HyD) with a bandpass filter of 605/70 nm. The Z-step size of the image (~200 nm) was automatically optimized by the system. Z-stacks were rendered in 3D in Leica LAS X software and analyzed in Fiji Image J. We confirmed that the dye preferentially stained minerals by comparing fluorescence images without and with minerals (**Figure S9**). To calculate the mineral thickness and fiber diameter, we did five cut section views along the field of view. Locations were sectioned randomly. We re-oriented the confocal view perpendicular to user. We utilized Image J measuring tool to manually measure the thickness of the mineral. In addition, we measured the hole diameter assuming it was a spaced occupied by the PVDF fiber. From each section cut, we averaged the fiber diameter and mineral thickness three times. The result summarized in Figure 3j showed good agreement with the data from the SEM analysis in Figure S7c.

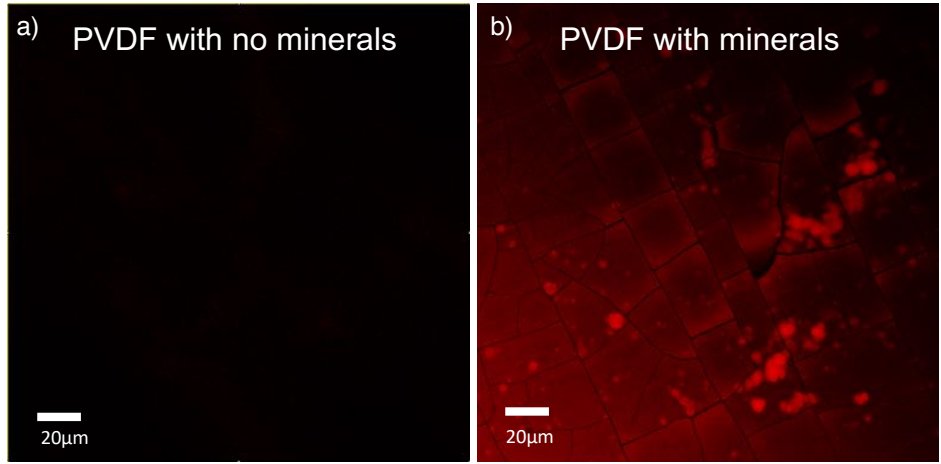

**Figure S9.** Staining process with Alizarin Red S. a) PVDF film without formed minerals showing no red fluorescence in the polymer. b) Mineralized PVDF film showing the expected red fluorescence in regions with formed minerals.

**Supporting Information S13: Analytical Model**

We further analyzed the data from the functionally graded material experiment (Supporting Information S7) to find the relation between the mineral height and the charge along the beam. Assuming small deformation, linear elastic response of a material and a linear relation between force and charge, we can obtain Eqnuation (S6):

$$Q = d_{33}F$$

$$Q = d_{33}(\sigma_{33}A) \tag{S6}$$

where $Q$ is charge (unit: C), $d_{33}$ is piezoelectric constant (unit: pC/N), $\sigma$ is stress (unit: MPa), and $A$ is area perpendicular to stress direction (unit: mm$^2$).[12]

From the stress distribution along the beam length in the **Figure S10**, the stress along the z-direction ($\sigma_{33}$) at a given location ($x$) can be described as Equation (S7) after a polynomial regression of second order (R$^2$=0.99), (see Figure S10b):

$$\sigma_{33} = 0.1x^2 - 4.1866x + 37.183 \tag{S7}$$

From the mineral formation distribution along the beam length, the mineral height at a location can be approximated as Equation. (S8) after a linear regression ($R^2=0.82$), (see Figure S10c):

$$MH = -1.52\times10^{-3}\, x + 1.79\times10^{-2} \tag{S8}$$

where $MH$ is mineral height (mm), $x$ is the beam length (mm). After solving for $x$ in Equation (S8), and replacing it into Equation (S7), and then into Equation (S6), we can obtain Equation. (S9):

$$Q \cong d_{33}A(4.33\times 10^4 MH^2) \tag{S9}$$

Then, the Equation (S9) can be rewritten as Equation (S10):

$$MH = \sqrt{\frac{Q}{4.33\times 10^4 d_{33} A}} \tag{S10}$$

Because charge ($Q$) is linearly proportional to stress ($\sigma$) (Equation S6), the mineral height ($MH$) is proportional to $Q^{1/2}$ or $\sigma^{1/2}$.

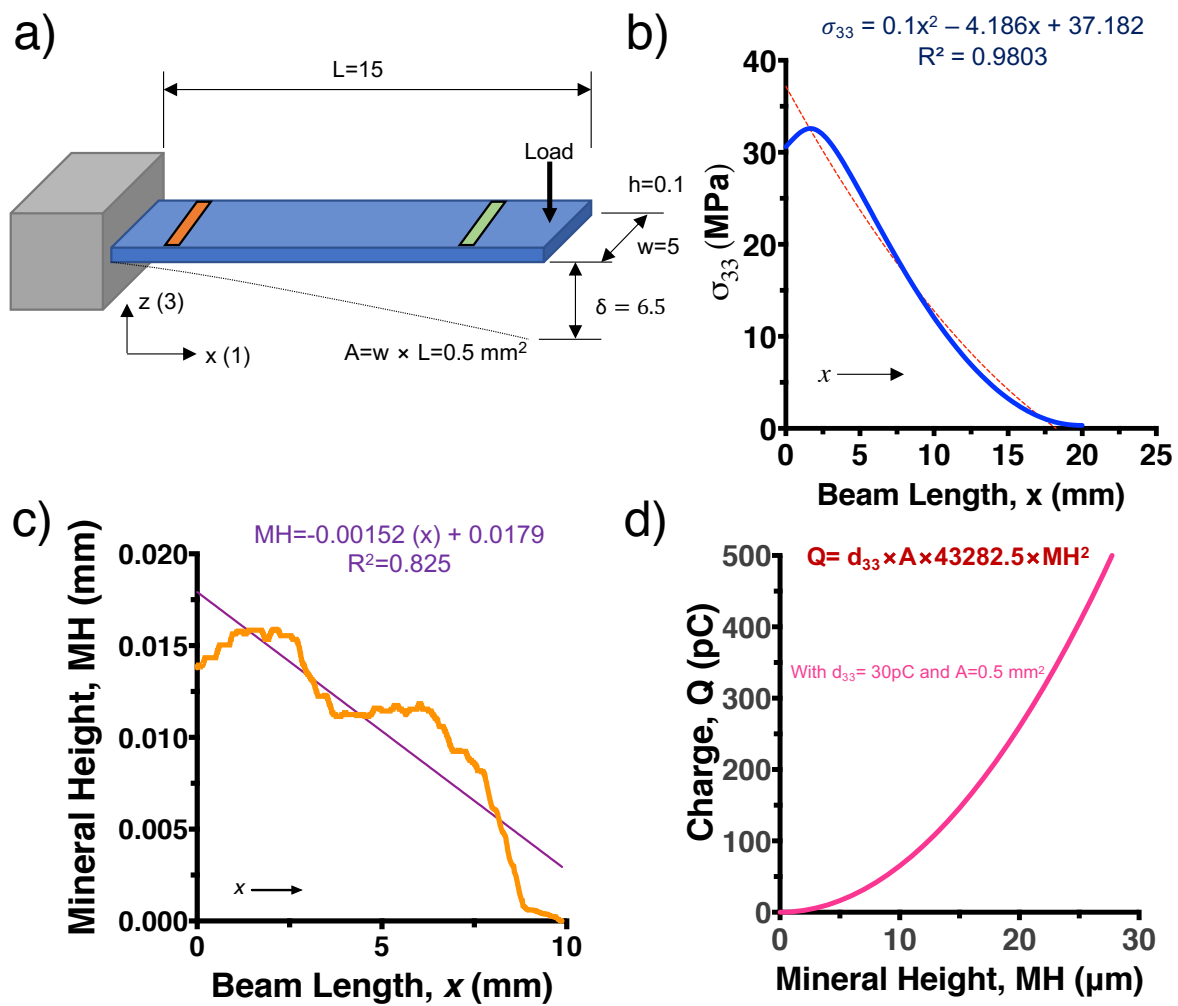

**Figure S10.** Analytical model. a) Cantilever configuration for a graded material experiment. b) Stress distribution along the beam length obtained from numerical model (Supporting Information S9). Higher stress at regions closer to the foundation. c) Mineral height along the beam length obtained from experiments (Supporting Information S7). Higher mineral deposited in regions of high stress of the cantilever beam. d) Analytical model obtained from experimental results in combination with numerical model and analytical equation describing the relationship between mineral height and electrical charge. Units in mm.